**Comment on "Coherent Electron Cooling"**

An interesting method proposed in [1] - coherent electron cooling (CEC) of ion beams in a storage ring of the Large Hadron Collider (LHC). The authors estimated that theirs method, in contrast to the other traditional methods of cooling, is an efficient cooling procedure for the proton beam at high energy. As the subject of beam cooling is important for practical applications, we decided to compare the method described in [1] with the one described earlier in [2].

In the method proposed in [1], a uniform cylindrical electron and ion beams pass along the modulatory section with the same speed. Heavy ions attract the electrons so that at the exit of the modulator section, the electron beam has enhanced charge density around each ion. Sequentially, the electron beam is separated from the ion one and enters the undulator, where each region with its high charge density emits a coherent undulator radiation (CUR). Emitted along the beam axis, CUR, passing through the front part of the moving beam, modulates its density and, in turn, this density modulation radiates a CUR superimposed on the previously emitted wave. In this case CUR becomes amplified as well as the density modulation of electron beam. This section is essentially a free-electron laser (FEL) in a mode of self-amplified stimulated emission (SASE). As a result, at the output of the FEL in the direction of its axis CUR has the form of a wave packet (wavelet) with length $K\lambda_{1,\min}$ and the electron beam density modulation at the same length, where $\lambda_{1,\min} \cong \lambda_u (1+ <p_\perp^2>)/2\gamma^2$ - is the wavelength of CUR emitted along the axis at the first harmonic, $K$ – is the number of periods in undulator, $\lambda_u$ - is the undulator period, $p_\perp = \beta_\perp \gamma$ is the undulator deflection parameter, $\beta_\perp = v_\perp /c$, $v_\perp$ - is the transverse component of the velocity, $\gamma = 1/\sqrt{1-\beta^2}$, $\beta = v/c$, $v$ -is a velocity of particle, $c$ - is the speed of light [3]. Further on the electron beam enters a kicker, where it merges with the ion beam. Electron beam keeps the static modulation of density in a form of the electron wave packets with the length $\beta K \lambda_{1,\min}$. In the kicker region, at the areas of periodic wave packets, the static accelerating electric field along the axis of the electron beam is relatively small (if compared with the transverse ones). In this field, the ions could be accelerated or slowed down. For the ion beam cooling the trajectories of electron and ion beams should be chosen so that the ions with zero transverse momentum and longitudinal velocity equal to the speed of the electron beam fall into their own electron wave packet to the point where the longitudinal and transverse electric field strengths are zeros. The ions with larger longitudinal momentum should fall into the decelerating region, and the ions with smaller longitudinal momentum - in the accelerating field of electron beam. The rate of cooling of ion beam is determined by the longitudinal component of electric field in the electron wave packets moving in the static fields of electron beam (in its coordinate system).

Below, aside from this dynamical model, we found the maximal possible strength of electromagnetic field excited in the electron beam based on more general considerations. We came to the conclusion that the rate of cooling of the ion beam in the LHC with this scheme [1] is about the same as the rate of enhanced optical cooling (EOC) and the fast laser cooling by the broad band laser beam, discussed in our earlier works about cooling of ions in the LHC [2], [3], [4]. The limiting rate of cooling in the CEC scheme itself looks to be higher than it was estimated in [1].

Let us consider the process in a system of coordinates, which is moving at the average speed of particles in the electron beam. In this system the modulated by density parts of the beam looks as a sequence of $K$ charged round discs with radius $R$ located along theirs common axis at the distance $h = \beta\gamma\lambda_{1,\min}$. The thickness of discs is much less than the distance between them so it could be neglected. The disc could be described by the surface charge density $\sigma$. In cylindrical system of coordinates ($r, z, \varphi$), which is moving with the speed of electron beam, the charge is $dq = \sigma r dr d\varphi$. The longitudinal component of electric field created by this element of charged disc is $dE_z = a dq /(r^2 + a^2)^{3/2}$, which for the whole disk comes to $E_z = 2\pi\sigma(1 - a/\sqrt{R^2 + a^2})$, where $a$ is the distance between disc and the reference point located at the axis $z$. For the disc located at the beginning of moving system of coordinates, the dependence of the field strength looks like

$$E_z|_{z<0} = -2\pi\sigma(1 + z/\sqrt{R^2 + z^2}), \qquad E_z|_{z>0} = 2\pi\sigma(1 - z/\sqrt{R^2 + z^2}). \tag{1}$$

For the neighboring disc located at the point $z = h$ the dependence of electrical field strength from coordinates looks like

$$E_z|_{z<h} = -2\pi\sigma[1 + (z-h)/\sqrt{R^2 + (z-h)^2}], \qquad E_z|_{z>h} = 2\pi\sigma[1 - (z-h)/\sqrt{R^2 + (z-h)^2}]. \tag{2}$$

Summarized electrical field strength between two discs at the axis is as the following

$$E_z\big|_{0<z<h} \cong -2\pi\sigma[(z-h)/\sqrt{R^2 + (z-h)^2} + z/\sqrt{R^2 + z^2}]\big|_{h\ll R} \cong -4\pi\sigma(z - h/2)/R,$$

$$E_z\big|_{z>h} \cong 2\pi\sigma[2 - (z-h)/\sqrt{R^2 + (z-h)^2} - z/\sqrt{R^2 + z^2}] \tag{3}$$

If all discs are shifted by $-h/2$, then with a new variable $\tilde{z} = z - h/2$ the first equation in (3) will look like the following

$$E_z\big|_{|\tilde{z}|<h/2} \cong -2\pi\sigma[(\tilde{z} - h/2)/\sqrt{R^2 + (\tilde{z} - h/2)^2} + (\tilde{z} + h/2)/\sqrt{R^2 + (\tilde{z} + h/2)^2}]\big|_{h\ll R} \cong -4\pi\sigma\tilde{z}/R \tag{4}$$

Equations (3) and (4) yield, that within approximation considered in the first order of $h/2R \ll 1$, the electrical field strength between discs $0 < z < h$ ($|\tilde{z}| < h/2$) is a linear function of $z$, drops from the value $E_z\big|_{z=0} \cong 2\pi\sigma h/R$ to $E_z\big|_{z=h} \cong -2\pi\sigma h/R$ and in the region $|\tilde{z}| < h/2$ does not depend on the distance between the discs $h$. Outside the region between discs $|\tilde{z}| > h/2$ the $E_z$ value is substantially higher and drops from $4\pi\sigma$ to zero value at $|\tilde{z}| > h/2$.

If $N_p$ pairs of discs located symmetrically along axis $\tilde{z}$ around the point with coordinate $\tilde{z} = 0$ with the distance between pairs $nh$, where $n = 1, 2, 3...$ - integer number, then the summarized electrical field strength between the first pair of discs will be defined by expression (see Appendix)

$$E_z\big|_{|\tilde{z}|<h/2}^{n>n_c} = (4\pi\sigma/h)[\sqrt{R^2 + [(\tilde{z} - N_p h/2)]^2} - \sqrt{R^2 + [(\tilde{z} + N_p h/2)]^2}]\big|_{n_c \gg 1} = -8\pi\sigma\tilde{z}/h \tag{5}$$

where $n_c = 2R/h$. The value $\sigma = eN_e/\pi R^2$, where the number of electrons in a disc $N_e = I \cdot \lambda_{1,\min} / \beta ec \cong 2 \cdot 10^8 \cdot \lambda_{1,\min}[cm] \cdot I[A]$, $I$ - the beam current in the laboratory coordinate system (the number of electrons in a disc, $\sigma$, the longitudinal component of the electric field strength and the radius of disk are invariants).

Let us attract attention, that expression (5) for the electrical field strength produced by the $N_p$ pairs of discs is valid in narrow margins $-h/2 < \tilde{z} < h/2$ (in in a region which is common for all pairs). In the neighboring region, $-3h/2 < \tilde{z} < -h/2$, this will be equivalent to a shift of the system of coordinate on the distance $\Delta\tilde{z} = -h$, reduced the number of pairs of discs down to the $N_p - 1$ and with addition of two discs at the opposite side. In this case the electrical field strength created in the neighboring regions by residual $N_p - 1$ pairs of discs could be described by formula (5). The residual two discs for

$N_p < n_c$, can create additional parasitic field strength compared with the one described by (5). For the further shift, the parasitic electrical field strength becomes much higher, than the field inside first pair of discs described by (5), which might result of reduction of cooling rate. Fortunately, not modulated long tails of the beam on both sides of the electron wave packet negate this effect.

*Example*. In the reference [1] for cooling of proton beam having energy 7 TeV in LHC was proposed to use a) electron beam with the energy 3.821 GeV ($\gamma = 7477$), current 100 A, diameter of 1 *mm*, the beam duration of 50 *ps* and the revolution frequency of ions $f = 1.1 \cdot 10^4$, b) FEL with a helical undulator having the length 65 m, magnetic field strength 0.9877 T and period 5 cm, kicker length $K\lambda_u = 10$ *m*, the relative energy spread of the proton beam $\delta_\varepsilon = \Delta\varepsilon / \varepsilon = 2 \cdot 10^{-4}$, $\Delta\varepsilon = 1.4 \cdot 10^9$ eV.

In this case $< p_\perp^2 > = (H/H_c)^2 \cong 21.3$ ($H_c \cong 2\pi mc^2 / e\lambda_u \cong 2140[G]$), the CUR is emitted at the wavelength $\lambda_{1,\min} = 10^{-6}$ cm, $h \cong 7.5 \cdot 10^{-3} cm$, the number of electrons in a bunch, $N_e = 9 \cdot 10^3$, discs charge density $\sigma = 3.1 \cdot 10^{-4}$, the electrical field strength according to (5), $E_{\tilde{z}}|_{n \gg n_c} = -8\pi\sigma\tilde{z}/h|_{\tilde{z}=-h/2} = 3.9 \cdot 10^{-3}[E] = 1.17$ [V/cm]. The maximal rate of the proton energy change is $\partial\varepsilon / \partial\tilde{z} = eE_{\tilde{z}}|_{\tilde{z}=-h/2} = 1.17[eV/cm]$. The energy losses of proton in the LHC is $P = eE_z K\lambda_u = 1.17 \cdot 10^3$ $eV/pass = 1.3 \cdot 10^7 eV/\sec$, the damping time $\tau_e = \Delta\varepsilon / P \cong 1.1 \cdot 10^2$ sec. Unfortunately in [1] there is no data on the transverse dimensions of the electron beam, proton beam, energy spread, the length of the modulator and kicker but presented final result – damping time ~1 hour. Their result is ~30 times higher than our limiting one if we use for estimations our beam dimensions and the kicker length. This value~30 could be referred to the lesser degree of beam modulation in [1], the kicker length and the proton beam energy spread.

It should be further noted that in [1] considered too idealized scheme of modulation of the electron beam. Thus, the size of the "induced charge" in the electron beam equal to the charge of a single proton cannot be concentrated at the distance $\delta < \lambda_{1,\min} = 4.5 \cdot 10^{-7}$ *cm* and so to generate effectively CUR at this wavelength. In Example considered, the average distance between particles for the current 100 A in the laboratory frame is ~$10^{-4}$ cm (in the moving frame - much higher). The limiting value of the longitudinal electric field strength component (5) in the modulated electron beam does not depend on the ion's charge state in the CEC scheme.

We would like to underline that the rate of the EOC and fast laser cooling of ion beams by the wide band laser beam, discussed in our earlier works [2], [3], [4] dedicated to the process of ion cooling in the LHC is comparable with CEC and for some types of ions could be even higher. Our methods are simpler and less expensive than CEC, however.

E.G.Bessonov
FIAN, Moscow, Russia

A.A.Mikhailichenko
Cornell U., CLASSE, Ithaca, U.S.A.

___________________________________

*Appendix*

$$E_z\big|_{|\tilde{z}|<h/2} \cong -2\pi\sigma \sum_1^\infty [(\tilde{z}-nh/2)/\sqrt{R^2+(\tilde{z}-nh/2)^2} + (\tilde{z}+nh/2)/\sqrt{R^2+(\tilde{z}+nh/2)^2}] \cong$$

$$-2\pi\sigma \int_1^\infty [(\tilde{z}-nh/2)/\sqrt{R^2+(\tilde{z}-nh/2)^2} + (\tilde{z}+nh/2)/\sqrt{R^2+(\tilde{z}+nh/2)^2}]dn \cong$$

$$-(4\pi\sigma/h)[-\int_{\tilde{z}-h/2}^{\tilde{z}-N_ph/2} (x/\sqrt{R^2+x^2})dx + \int_{\tilde{z}+h/2}^{\tilde{z}+N_ph/2} (y/\sqrt{R^2+y^2})dy \cong$$

$$-(4\pi\sigma/h)[-\sqrt{R^2+x^2}\big|_{\tilde{z}-h/2}^{\tilde{z}-N_ph/2} + \sqrt{R^2+y^2}\big|_{\tilde{z}+h/2}^{\tilde{z}+N_ph/2} \cong (4\pi\sigma/h)[\sqrt{R^2+x^2}\big|_{\tilde{z}-h/2}^{\tilde{z}-N_ph/2} - \sqrt{R^2+y^2}\big|_{\tilde{z}+h/2}^{\tilde{z}+N_ph/2} \cong$$

$$(4\pi\sigma/h)[\sqrt{R^2+(\tilde{z}-N_ph/2)^2} - \sqrt{R^2+(\tilde{z}+N_ph/2)^2} + \sqrt{R^2+(\tilde{z}+h/2)^2} - \sqrt{R^2+(\tilde{z}-h/2)^2} \cong$$

$$(4\pi\sigma R/h)[\sqrt{1+[(\tilde{z}-N_ph/2)/R]^2} - \sqrt{1+[(\tilde{z}+N_ph/2)/R]^2} \cong$$

$$(4\pi\sigma R/h)[\sqrt{1+(\tilde{z}-N_p^2h^2/4)/R^2 - \tilde{z}N_ph/R^2} - \sqrt{1+(\tilde{z}^2+N_p^2h^2/4)/R^2 + \tilde{z}N_ph/R^2} \cong$$

$$(4\pi\sigma R/h)[\sqrt{1+N_p^2h^2/4R^2} \cdot [\sqrt{1-\tilde{z}(N_ph/R^2)/(1+N_p^2h^2/4R^2)} - \sqrt{1+\tilde{z}(N_ph/R^2)/(1+N_p^2h^2/4R^2)}] \cong$$

$$(4\pi\sigma R/h)(N_ph/2R)[\sqrt{1-4\tilde{z}/N_ph} - \sqrt{1+4\tilde{z}/N_ph} \cong -(4\pi\sigma R/h)(N_ph/2R)4\tilde{z}/N_ph \cong -8\pi\sigma\tilde{z}/h$$